%% file: main.tex
\documentclass[letterpaper, 10 pt, conference]{ieeeconf}  \IEEEoverridecommandlockouts                              
\usepackage{graphics} 
\usepackage{epsfig} 
\usepackage{mathptmx} 
\usepackage{times} 
\usepackage{amsmath} 
\usepackage{amssymb}  
\usepackage{xspace} 
\usepackage{authblk}
\usepackage{listings}
\usepackage{parskip}

\usepackage[utf8]{inputenc}
\usepackage[english]{babel}
\usepackage[backend=biber,sorting=none]{biblatex}
\usepackage [autostyle, english = american]{csquotes}
\MakeOuterQuote{"}

\usepackage{pgfplots}
\usepackage{filecontents}

\usepackage{fancyhdr}

\usepackage{algorithm}
\usepackage{algpseudocode}

\usepackage{cleveref}
\algnewcommand{\algorithmicgoto}{\textbf{go to}}%
\algnewcommand{\Goto}[1]{\algorithmicgoto~\ref{#1}}%
\algrenewcommand\textproc{}
\algnewcommand{\LeftComment}[1]{\Statex \(\triangleright\) #1}

\makeatletter
\newenvironment{breakablealgorithm}
  {
   \begin{center}
     \refstepcounter{algorithm}
     \hrule height.8pt depth0pt \kern2pt
     \renewcommand{\caption}[2][\relax]{
       {\raggedright\textbf{\ALG@name~\thealgorithm} ##2\par}%
       \ifx\relax##1\relax 
         \addcontentsline{loa}{algorithm}{\protect\numberline{\thealgorithm}##2}%
       \else 
         \addcontentsline{loa}{algorithm}{\protect\numberline{\thealgorithm}##1}%
       \fi
       \kern2pt\hrule\kern2pt
     }
  }{
     \kern2pt\hrule\relax
   \end{center}
  }
\makeatother

\bibliography{citations.bib}

\title{\LARGE \bf
GCList: Garbage Collection in Concurrent Sets
}

\author{Shekhar Bhandakkar, Jonathan Marbaniang and Sathya Peri
\\Department of Computer Science \& Engineering,
\\Indian Institute of Technology, Hyderabad, India
}
\affil{\textit {\{cs14btech11006, cs16mtech11006, sathya\_p\}@iith.ac.in}}

\input{macros}

\begin{document}

\maketitle
\thispagestyle{empty}
\pagestyle{empty}

\begin{abstract}
Garbage Collection in concurrent data structures, especially lock-free ones, pose multiple design and consistency challenges. In this instance, we consider the case of concurrent sets. A set is a collection of elements, where the elements are ordered and distinct. These two invariants are always maintained at every point in time.\\
Sets are usually represented as a linked list of nodes, with each node denoting an element in the Set. Operations on the set include adding elements to the set, removing elements from it and searching for elements in it. Currently, multiple implementations of concurrent sets already exist. LazyList\cite{lazy_list}, Hand-over-hand List\cite{hoh} and Harris' List\cite{lfree} are some of the well-known implementations. However none of these implementations employ, or are concerned with garbage collection of deleted nodes. Instead each implementation ignores deleted nodes or depends on the language's garbage collector to handle them.\\
Additionally, Garbage collection in concurrent lists, that use optimistic traversals or that are lock-free, is not trivial.\\
For example, in Lazy List and Harris' List, they allow a thread to traverse a node or a sequence of nodes after these nodes have already been removed from the list, and hence possibly deleted. If deleted nodes are to be reused, this will potentially lead to the ABA problem.\cite{queue_aba}\\
Moreover, some languages like C++ do not have an in-built garbage collector. Some constructs like Shared Pointers\cite{SPtr} provide a limited garbage collection facility, but it degrades performance by a large scale. Integrating Shared Pointers into a concurrent code is also not a trivial task.\\
In this paper, we propose a new representation of a concurrent set, GCList, which employs in-built garbage collection. We propose a novel garbage collection scheme that implements in-built memory reclamation whereby it reuses deleted nodes from the list. We propose both lock-based and lock-free implementations of GCList. The garbage collection scheme works in parallel with the Set operations.\\
In our experiments with varying workloads and randomised Set operations, GCList shows comparable performance to LazyList\cite{lazy_list} \& Harris' List\cite{lfree} while outperforming Shared Pointers\cite{SPtr}, Hazard Pointers\cite{HP} and Hand-over-hand List\cite{hoh}. GCList also consumed way lesser memory as compared to LazyList\cite{lazy_list} and Harris' List\cite{lfree} and is comparable to Shared Pointers\cite{SPtr} and Hazard Pointers\cite{HP}.
\end{abstract}


\input{intro}
\input{model}

\input{related_work}
\input{gc-algo}
\input{gc-pool}
\input{evaluation}
\input{conc}
\printbibliography
\input{appendix}

\end{document}

%% file: macros.tex
\newcommand{\punt}[1]{}
\newcommand{\cmnt}[1]{}
\newcommand{\ignore}[1]{}

\newcommand{\mth} {method\xspace}

\newcommand{\lble} {linearizabale\xspace}
\newcommand{\lbty} {linearizability\xspace}
\newcommand{\lp} {linearization point\xspace}

%% file: intro.tex
\section{Introduction}
List-based implementation of concurrent sets are fairly common. LazyList\cite{lazy_list}, Hand-over-Hand List\cite{hoh} and Harris's LockFreeList\cite{lfree} are some common examples. However none of these implementations address the issue of garbage collection of nodes deleted from the list. Either the algorithm ignores the issue or it relies on the language’s garbage collector to handle it for them.
\par
There are several reasons to implement our own memory management scheme. Languages such as C and C++ do not provide garbage collection and often it is more efficient to do our own memory management. C++ has some constructs like Shared Pointers\cite{SPtr} that offer limited garbage collection facility. Other garbage collection techniques like Stop-the-World are also available. Even though Shared Pointers, Hazard Pointers\cite{HP} and these other garbage collection schemes are very generic techniques, since they can be applied to almost all concurrent data structures, they are expensive and cost a lot in terms of performance and the extra data structures required to implement them.
\par
Integrating Shared Pointers, Hazard Pointers and these other garbage collection schemes into a concurrent data structure is also not a trivial task. And more often than not, they are not very optimized for performance. They become even more complicated in case of lock-free data structures employing lock-free methods. Garbage collection, in these cases, is byzantine.\cite{queue_aba}
\par
In this paper, we concentrate on the garbage collection scheme for a concurrent set. We introduce a new representation of a concurrent set, \textbf{GCList}, with in-built garbage collection. Nodes that are removed from the set are collected in a "Pool" of deleted nodes, to be reused for later add operations. We introduce both lock-based and lock-free versions of GCList. We use the terms node, key and value interchangeably in this paper.

%% file: model.tex
\section{System Model \& Preliminaries}
\label{sec:System-Model-Preliminaries}
\vspace{-2mm}
In this paper, we assume that our system consists of finite set of $p$ processors, accessed by a finite set of $n$ threads that run in a completely asynchronous manner and communicate using shared objects. The threads communicate with each other by invoking higher-level \mth{s} on the shared objects and getting corresponding responses. Consequently, we make no assumption about the relative speeds of the threads. We also assume that none of these processors and threads fail. \\

\noindent
\textbf{Safety:} To prove a concurrent data structure to be correct, \textit{\lbty} proposed by Herlihy \& Wing \cite{HerlWing:1990:TPLS} is the standard correctness criterion in the concurrent world. They consider a history generated by a data structure which is collection of \mth invocation and response events. Each invocation of a method call has a subsequent response. A history is \lble if it is possible to assign an atomic event as a \emph{\lp} inside the execution interval of each \mth such that the result of each of these \mth{s} is the same as it would be in a sequential history in which the \mth{s} are ordered by their \lp{s} \cite{HerlWing:1990:TPLS}. \\

\ignore{
\textbf{Linearizability:} To prove a concurrent data structure to be correct, \textit{\lbty} proposed by Herlihy \& Wing \cite{HerlWing:1990:TPLS} is the standard correctness criterion in the concurrent world. They consider a history generated by a data structure which is collection of \mth invocation and response events. Each invocation of a method call has a subsequent response. A history to be \lble if (1) The invocation and response events can be reordered to get a valid sequential history. (2) The generated sequential history satisfies the object's sequential specification. (3) If a response event precedes an invocation event in the original history, then this should be preserved in the sequential reordering.  \\
}

\noindent
\textbf{Progress:} The \emph{progress} properties specifies when a thread invoking \mth{s} on shared objects completes in presence of other concurrent threads. Some progress conditions used in this paper are mentioned here which are based on the definitions in Herlihy \& Shavit. The progress condition of a method in concurrent object is defined as: (1) Blocking: In this, an unexpected delay by any thread (say, one holding a lock) can prevent other threads from making progress. (2) Deadlock-Free: This is a \textbf{blocking} condition which ensures that \textbf{some} thread (among other threads in the system) waiting to get a response to a \mth invocation will eventually receive it. (3) Wait-Free: This is a \textbf{non-blocking} condition which ensures that \textbf{every} thread trying to get a response to a \mth, eventually receives it\cite{Herlihy:WFS:TPLS:1991}.

%% file: related_work.tex
\section{Related Work}
We discuss some of the list-based set algorithms in this section and some existing garbage collection techniques that can be used in concurrent sets.

\subsection{Hand-Over-Hand List}
In this list-based representation of a set, also called \textbf{lock-coupling}\cite{hoh}, each thread traverses the list from the head of the list, while acquiring fine-grained locks in a hand-over-hand manner. Each thread acquires the lock for the next node and then releases the lock for the current node.\\
All operations require the usage of locks which may affect the overall performance of the list, even though garbage collection in this list is a fairly trivial task.

\subsection{LazyList}
An improvement over the Hand-over-Hand list is the LazyList\cite{lazy_list}. Threads traverse the list \textbf{optimistically}, without using any locks. Nodes are locked only when the required pair are found. An additional boolean field called "marked" field is associated with every node. The "marked" field is used to identify nodes that have been deleted but are still reachable from the head of the list.
\par
In LazyList, nodes are deleted in two steps:\\
- \textbf{Logical deletion:} The marked field is set to true.\\
- \textbf{Physical deletion:} The node's predecessor's next reference is swung to the node's successor.
\par
The contains method is completely wait-free. It traverses the list without using any locks. It's easy to see that garbage collection, in this case, is not so trivial. It may lead to an issue known as the "\textbf{ABA Problem}"\cite{queue_aba}.

\begin{figure}[h!]
\centering
\includegraphics[scale=0.25]{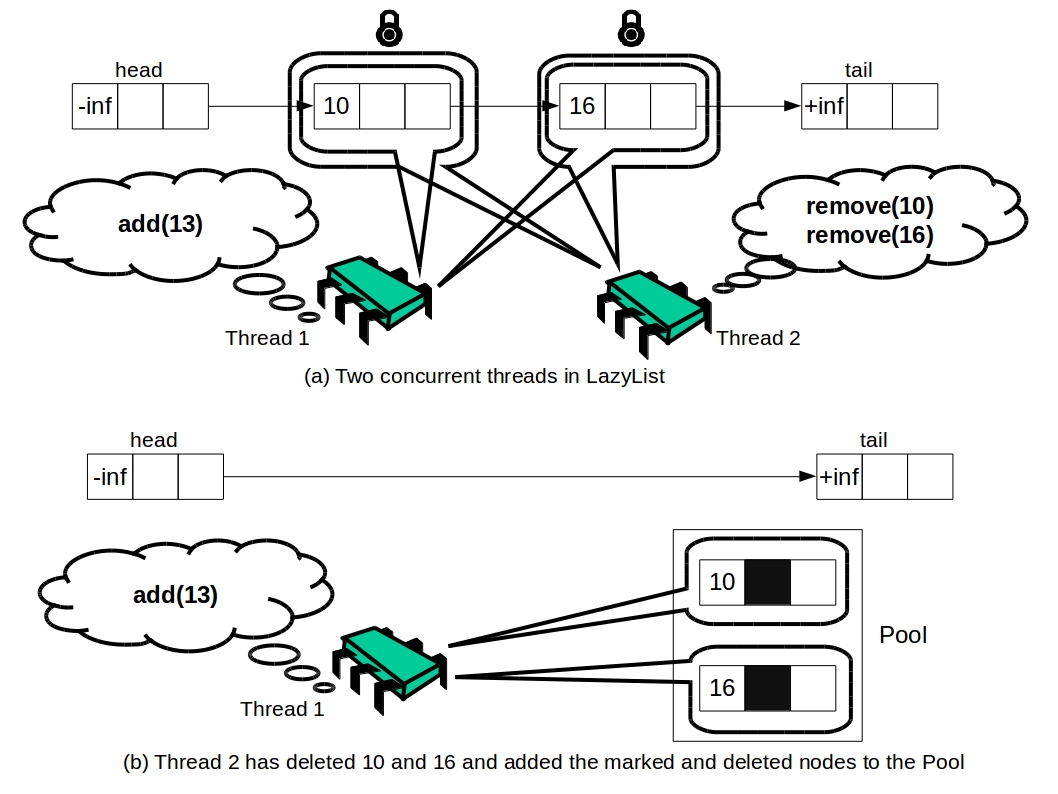}
\caption{The ABA Problem in LazyList (Part 1)}
\end{figure}

\begin{figure}[h!]
\centering
\includegraphics[scale=0.25]{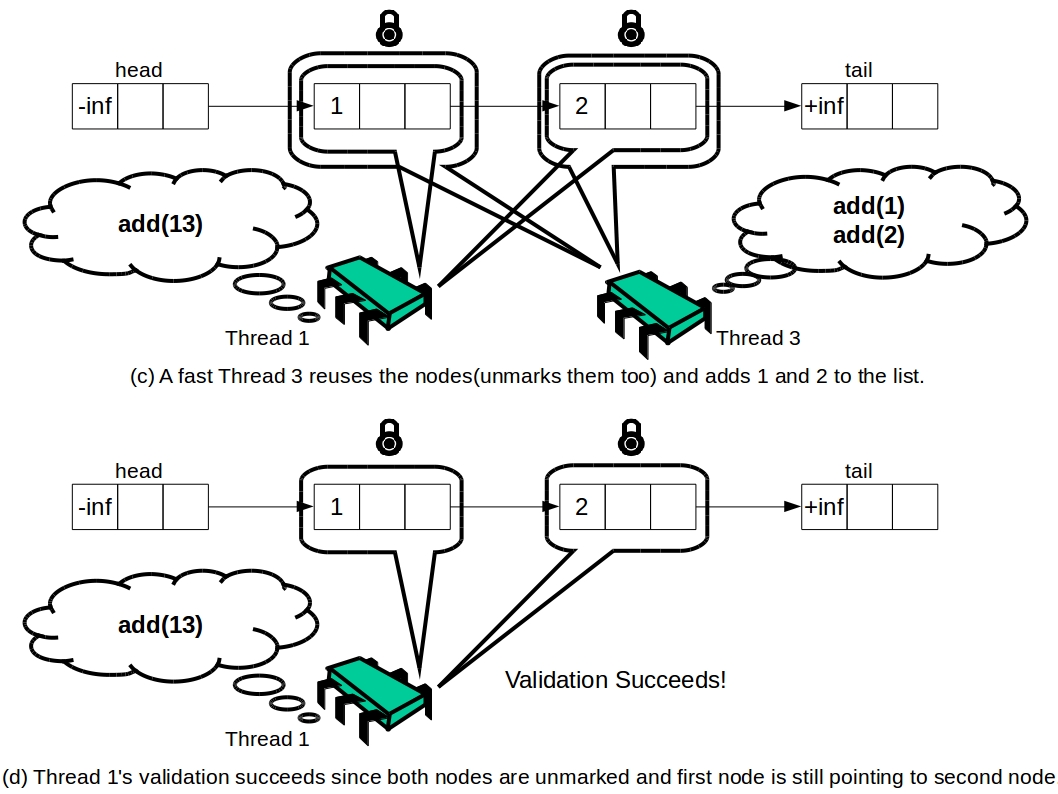}
\caption{The ABA Problem in LazyList (Part 2)}
\end{figure}

\begin{figure}[h!]
\centering
\includegraphics[scale=0.25]{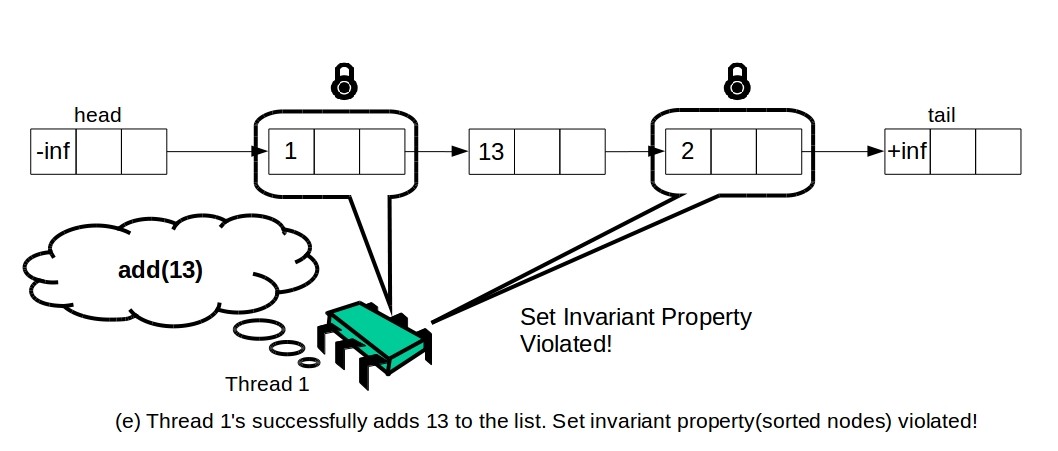}
\caption{The ABA Problem in LazyList (Part 3)}
\end{figure}

\subsection{LockFreeList}
The LockFreeList\cite{HerlihyShavit:AMP:Book:2012} is an extension of the LazyList\cite{lazy_list}, where locks are eliminated altogether from the list operations and all the methods are non-blocking\cite{Herlihy:WFS:TPLS:1991}.

\par
The list uses an AtomicMarkableReference\cite{AMR} object as a part of it's structure, which allows a thread to atomically read and update both the boolean mark and the next reference of a node. The list also uses compareAndSet or CAS calls for its operations.

\par
The remove method is similar to LazyList\cite{lazy_list}, in that deletion is done in two steps.\\
- A CAS call is used to set the marked field of a node.\\
- Another CAS call is used on the node's predecessor to physically delete the node from the list.

\par
An important difference between LockFreeList\cite{HerlihyShavit:AMP:Book:2012} and LazyList\cite{lazy_list} is that LockFreeList never traverses logically marked nodes. Instead the encountered marked nodes are physically deleted from the list. Essentially, threads "help" out other slower threads that have completed the first CAS call but not the second.

\par
It can also be seen that similar to LazyList\cite{lazy_list}, LockFreeList\cite{HerlihyShavit:AMP:Book:2012} is also vulnerable to the ABA problem\cite{queue_aba}.

\subsection{Reference Counting}
In a dynamic and concurrent data structure, arbitrary objects can continuously
and concurrently be added or removed from the data structure. And multiple owners may have a reference to the shared objects. Unsafe freeing of a node may lead to safety issues and possible crashes\cite{RefCount}.\\
So, before freeing a shared object, it should be checked that there are no remaining references to it. This should also include possible local references to the shared object that any thread might have, as a read or write access to the memory of a reclaimed object might be fatal to the correctness of the data structure andor to the whole system\cite{RefCount}.

\par
In the “Reference-Counting” category of garbage collection techniques, shared counters are assigned to objects and they are used to count the number of references to any object at any given time\cite{RefCount}. In other words, a group of owners share the ownership for an object.This group is responsible for deleting that object when the last one among them releases that ownership. The shared object can be freed if and only if the counter becomes zero\cite{RefCount2}.

\par
This method, however is expensive. A shared atomic counter has to be associated with every object\cite{RefCount}\cite{RefCount2}. Getting a reference to an object and incrementing the shared counter has to be an atomic operation. Same thing applies when losing the reference to the object and decrementing the shared counter. Even a simple read operation from the shared object has to increment the shared counter. Essentially, the memory read becomes a read-modify-write operation\cite{ThrScan}.

\subsection{Pointer-based techniques}
Pointer-based techniques such as Hazard Pointers\cite{HP} explicitly mark live objects (objects that threads can access) which are not de-allocated. Pointer-based schemes suffer from two limitations: they must be customized to the data structure at hand, which makes them difficult to deploy; they publish each pointer that is used in a shared memory location, which is expensive in terms of synchronization.

\par
Hazard Pointers (HP) and other pointer-based techniques will typically publish the pointer to each object they use, and then check that the pointer has not changed in the meantime. Such approach guarantees that an object which has been deleted will not be later dereferenced, at the cost of each reader doing synchronization on a per-object basis.

\par
Because it requires validation of the pointer that will be accessed next, Hazard Pointers are lock-free for readers, although in some situations they can be made wait-free for readers. HP is wait-free bounded for reclamation, with the bound being proportional to the number of threads times the number of hazard pointers, because each reclaimer has to scan all the hazard pointers of all the other threads before deleting a node. In HP the retired nodes are placed in a retired list which is scanned once its size reaches an R threshold. In terms of memory usage, when the R factor is set to the lowest setting of 1, each reclaimer can have at most a list of retired nodes with a size equal to the number of threads minus 1, times the number of hazard pointers. If each thread has one such list of nodes pending to be deleted, at any given point in time there are at most O($N_{threads}^2$) nodes to be deleted.

%% file: gc-algo.tex
\section{Our Algorithm: GCList}
This paper introduces the \textbf{GCList}, a list-based set algorithm with an in-built garbage collection scheme. The set is represented as a linked-list of nodes, supporting the following operations:\\
- \textbf{add(key)}, adds key to the set, and returns true if and only if key was not already present in the set.\\
- \textbf{remove(key)}, removes key from the set, and returns true if and only if key was present in the set.\\
- \textbf{contains(key)}, searches for key in the set, and returns true if and only if key is present in the set.\\
We introduce two versions of GCList, a blocking version or \textbf{GCLBList} and a non-blocking version or \textbf{GCLFList}.
\par
The pseudo-code for both the versions has been kept in the appendix.

\subsection{GCLBList}
Each node in the list consists of three fields: the key field, an \textbf{AtomicStampedReference}\cite{ASR} object called as infoNext and a lock associated with the node. We have implemented our own AtomicStampedReference\cite{ASR} in C++. The list is ordered according to the keys of each node. infoNext contains a reference to the next node in the list and an integer stamp associated with the node. Both the stamp and the reference can be read and updated atomically\cite{ASR}. The lock field is a lock used for synchronization.
\begin{lstlisting}[language=C++,caption=GCLBList Node,captionpos=b]
class Node
{
    int key;
    AtomicStampedReference<Node> infoNext;
    mutex nodeLock;
};
\end{lstlisting}

\begin{figure}[h!]
\centering
\includegraphics[scale=0.4]{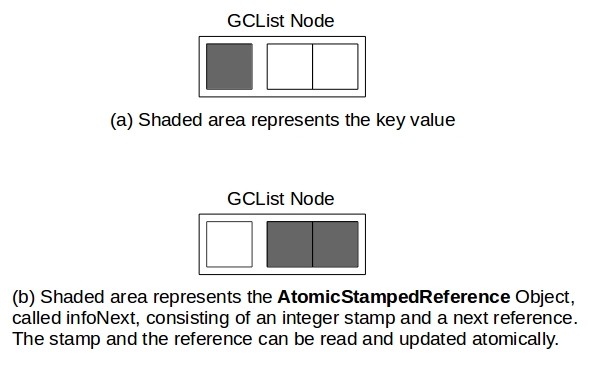}
\caption{GCList Node and it's components}
\end{figure}

As mentioned earlier, we consider three operations on the list i.e. add, remove and contains. However we factor out functionality common to the add and remove methods by creating an inner Window class to help navigation. The common functionality is used to optimistically traverse the list and “find” the required pair of nodes required for each operation. The find method then returns the references to the nodes and their respective stamps in a Window object to the calling method.

\subsubsection{The find method}
The find method is used by the add and remove methods to optimistically traverse the list. The thread gets a reference to the “head” node and keeps traversing the list in an optimistic hand-over-hand fashion. At every step of the traversal, the infoNext's reference and stamp fields of a node are read atomically\cite{ASR}. The thread keeps traversing the list until it finds the relevant pair of nodes, pred and curr. curr holds a reference to the first node with a key greater than or equal to the key that is being searched, in the list, with pred being curr's predecessor. The find method returns a window object, containing references to pred and curr along with their respective stamps, to the calling method.
\par
An important observation to be made here is the use of stamps during traversal. Stamps are used to detect synchronization conflicts by a traversing thread. This can be inferred from the working of the remove method later. If at any time during a thread's traversal, the stamp of the pred node changes, a synchronization conflict with another “removing” thread is detected. The current thread “retries” it's traversal from the head node.

\subsubsection{The validate method}
The validate method is used to ensure that the calling method has locked the correct pair of nodes. It uses the stamps and references returned by the find method to ensure that both pred and curr are still present in the list and pred is still pointing to curr. If the stamps of either node has changed or pred is no longer pointing to curr, then it signifies a synchronization conflict with another thread. The current thread then restarts it's execution.

\subsubsection{The remove method}
The remove method is used to remove key from the set, returning true if and only if key was in the set. It calls the “find” method to determine the correct pair of nodes for the remove operation. The nodes are locked and then validation is performed using the “validate” method. If validation fails, the nodes are unlocked and the thread retries, otherwise it continues it's operation.
\par
Deletion is performed in two steps:\\
- \textbf{Step 1:} pred's infoNext's reference is swung to curr's infoNext's reference and pred's infoNext's stamp is incremented by one. This operation to update pred's infoNext's reference and stamp fields is atomic.\\
- \textbf{Step 2:} curr's infoNext's stamp is incremented by 1. This marks the successful deletion of curr from the list.\\
Step 2 is the \textbf{\lp} for the remove method.
\par
After curr has been successfully deleted, it is added to the “Pool”. A \textbf{Pool} is a concurrent data structure which is used to hold the deleted nodes. These deleted nodes can now be reused for later add operations.
\par
Now, an important thing to discuss in this section is why does a thread traversing the list, in the find or contains method, has to retry if the pred's stamp changes. Based on the working of the “find” method, we can see that if any thread has a reference to curr, it should also have read pred's old stamp. This is because reads from an AtomicStampedReference\cite{ASR} object is atomic. At this point, if curr were to be deleted from the list, pred's stamp would have been incremented, in Step 1. Again, this updation of pred's infoNext fields is atomic.
\par
If the current thread were to continue it's traversal, it may instead traverse the Pool or some other part of the list, since we have no guarantees about curr's position after it's deletion. Instead, before advancing pred and curr in the list, we check pred's stamp again. If it has changed, it implies that curr may have been deleted and the current thread is in a synchronization conflict with a removing thread. The current thread then restarts it's traversal from the list's head again. If pred's stamp is unchanged though, it implies that curr is still a part of the list and the thread can advance pred and curr.
\par
Conversely, we can say that if a thread has read pred's updated stamp at the first read, then it cannot have a reference to curr. Again, this is because the updation of pred's infoNext fields is atomic\cite{ASR}.

\begin{figure}[h!]
\centering
\includegraphics[scale=0.25]{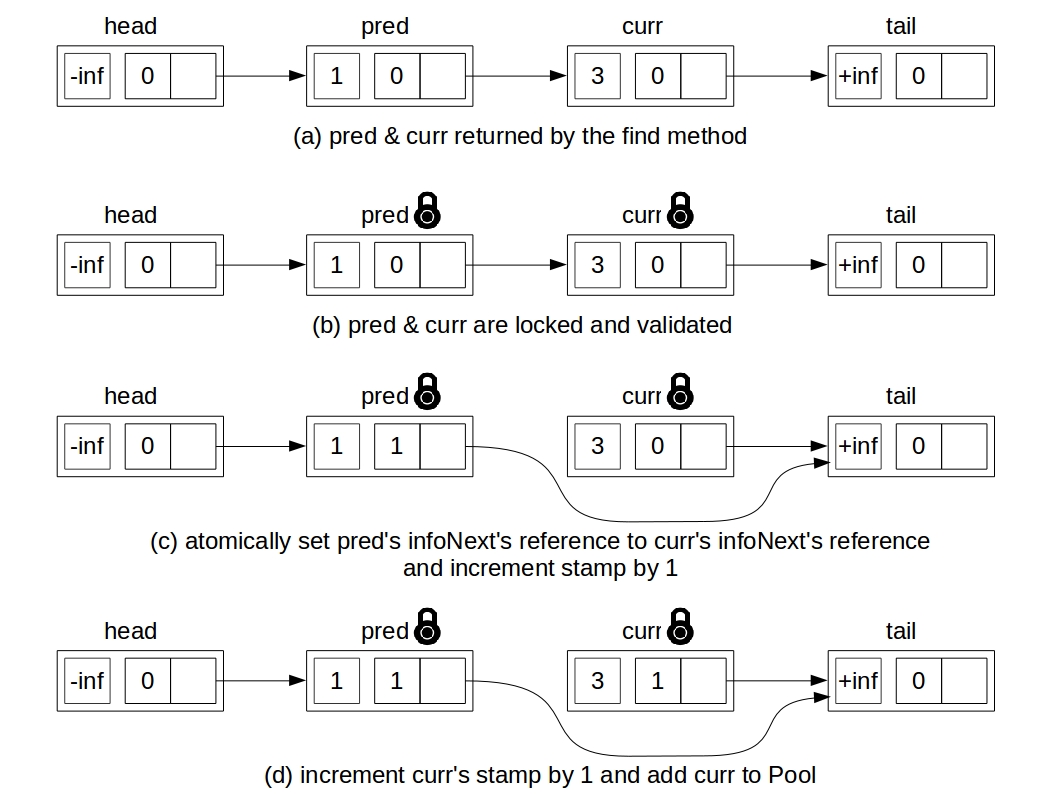}
\caption{GCLBList: Remove Steps}
\end{figure}

\begin{figure}[h!]
\centering
\includegraphics[scale=0.24]{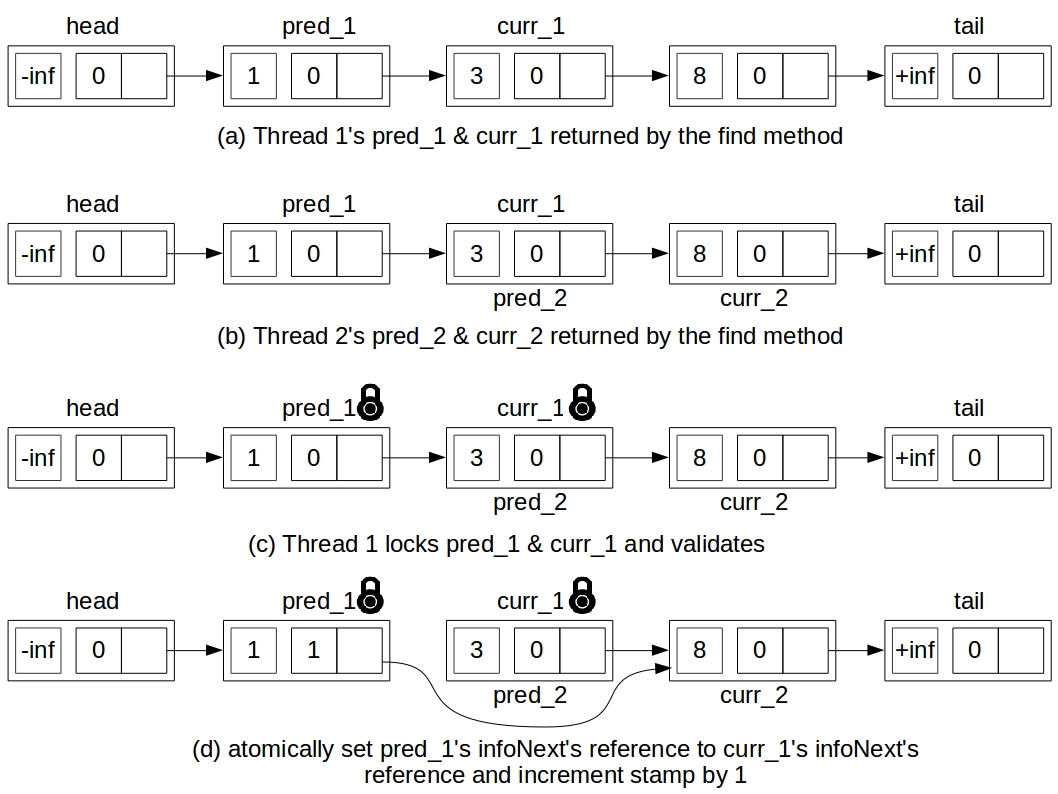}
\caption{GCLBList: Two concurrent removing Threads(Part 1)}
\end{figure}

\begin{figure}[h!]
\centering
\includegraphics[scale=0.24]{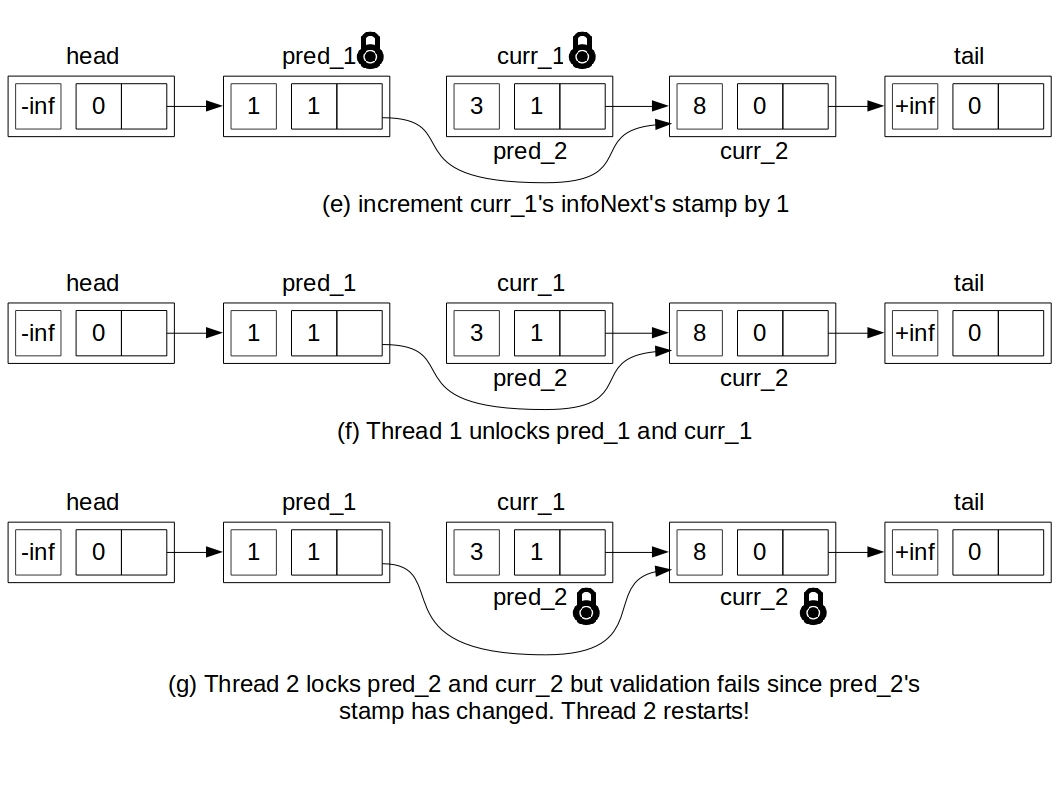}
\caption{GCLBList: Two concurrent removing Threads(Part 2)}
\end{figure}

\subsubsection{The add method}
The add method is used to add a key to the list if and only if the key is not already present in the list. It calls the “find” method to determine the correct pair of nodes for the add operation. The nodes are locked and then validation is performed using the “validate” method. If validation fails, the nodes are unlocked and the thread retries, otherwise it continues it's operation. The thread then queries the Pool(a data structure containing deleted nodes) for a node. If the Pool is not empty, a node is returned to be reused. Else, the thread creates a new node. It then inserts the new node, unlocks pred and curr and returns true.\\
The step in which pred's infoNext's reference is set to the new node is the \textbf{\lp} for the add method.

\begin{figure}[h!]
\centering
\includegraphics[scale=0.23]{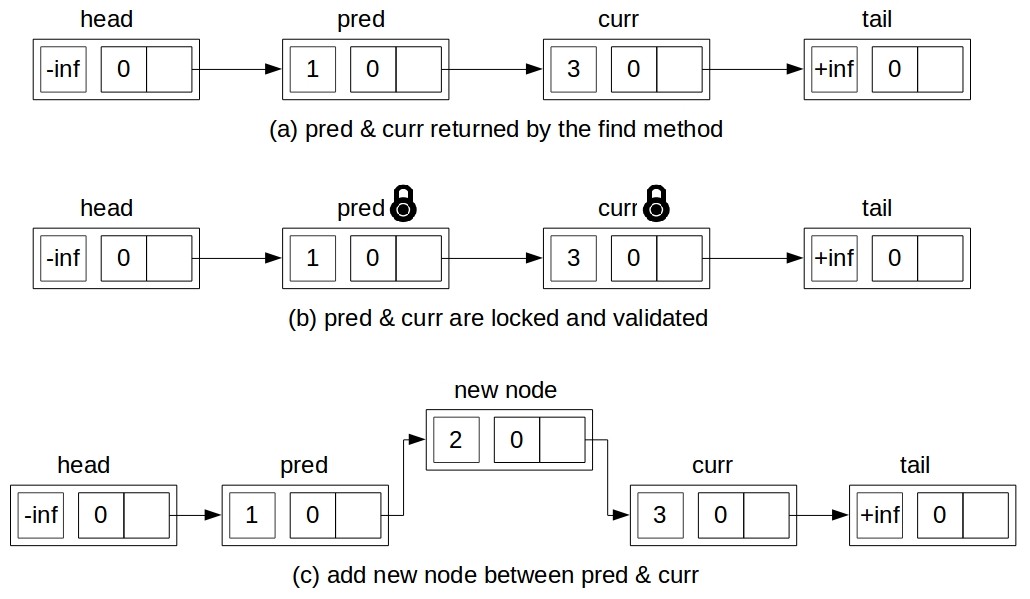}
\caption{GCLBList: Add Steps}
\end{figure}

\subsubsection{The contains method}
The contains method is similar to the find method. It starts from the “head” node and keeps traversing the list in an optimistic hand-over-hand fashion. At every step of the traversal, the infoNext's reference and stamp fields of a node are read atomically\cite{ASR}. The thread keeps traversing the list until it finds the first node with a key greater than or equal to the key that is being searched.
\par
Similar to the find method, stamps are used to detect synchronization conflicts during traversal. If at any time during a thread's traversal, the stamp of the pred node changes, a synchronization conflict with another “removing” thread is detected. The current thread “retries” it's traversal from the head node.
\par
The method returns true if and only if the key is present in the list. A successful contains is \textbf{linearized} when a matching key is found and the stamp of the predecessor hasn't changed from it's previous value.

\subsection{GCLFList}
GCLFList is the non-blocking version of our list-based set algorithm.
\par
Each node in the list now consists of two fields, the key field and an AtomicStampedReference\cite{ASR} object called as infoNext. The list is ordered according to the keys of each node. infoNext contains a reference to the next node in the list and an integer stamp associated with the node. Both the stamp and the reference can be read and updated atomically\cite{ASR}. There is no lock field associated with the node anymore.
\begin{lstlisting}[language=C++,float,floatplacement=H,caption=GCLFList Node,captionpos=b]
class Node
{
public:
	int key;
	AtomicStampedReference<Node> infoNext;
}
\end{lstlisting}
\par
We instead use atomic functions like compareAndSet\cite{ASR} or CAS to perform our operations on the list. Atomic operations\cite{ASR} are used to atomically read and update the AtomicStampedReference object associated with each node. However, this also leads to complications. For example, if we follow the deletion steps of GCLBList, what happens in the case of two adjacent concurrent remove operations, using CAS? We can see that one of the nodes won't be removed from the list.
\par
To solve this problem, we need a way to identify a marked node in the list, even though it may still be present in the list i.e. a logically deleted node. We differentiate between a logically deleted node and a node that is a part of the list by using parity of stamp.\\
- A node with an \textbf{even stamp} is a part of the list.\\
- A node with an \textbf{odd stamp} denotes a node that has been deleted from the list.
\par
The deletion operation is also divided into two steps\\
- \textbf{Logical Deletion:} Increment curr's stamp by 1 using CAS\cite{ASR} i.e marking curr. This step is the \textbf{\lp} of the remove method.\\
- \textbf{Physical Deletion:} Swing pred's infoNext's reference to curr's infoNext's reference and increment pred's infoNext's stamp by 2, atomically using CAS\cite{ASR}.
\par
We also adopt the concept of \textbf{Helping} i.e. if a traversing thread encounters a logically deleted or marked node, it attempts to first remove the node from the list, before advancing forward.

\subsubsection{The find method}
The find method is used by the add and remove methods to optimistically traverse the list. The thread gets a reference to the “head” node and keeps traversing the list in an optimistic hand-over-hand fashion. At every step of the traversal, the next reference and stamp of a node is read atomically\cite{ASR}. The thread keeps traversing the list until it finds the relevant pair of nodes, pred and curr. It returns a window object, containing references to pred and curr along with their respective stamps, to the calling method.
\par
As mentioned above, each time the thread encounters a marked node i.e. a node with an odd stamp, it attempts to physically delete the node first before advancing. If the CAS operation for the physical deletion succeeds, the node advances forward. Else it retries. Threads never traverse marked nodes because they lead to consistency issues.
\par

For example, find may return a marked pred and an unmarked curr to the remove method trying to add a new node between pred and curr. If pred is physically removed by another thread before the new node could be added, the new node would end up being not added to the list. This difficulty arises because the current thread is not holding locks on pred and curr.
\par
Similar to the previous find method, stamps are also used to detect synchronization conflicts by a traversing thread. If at any time during a thread's traversal, the stamp of the pred node changes, a synchronization conflict with another “removing” thread is detected. The current thread “retries” it's traversal from the head node.

\subsubsection{The remove method}
The remove method is used to remove key from the set, returning true if and only if key was in the set.  It calls the “find” method to determine the correct pair of nodes for the remove operation.\\
Deletion of “curr” is performed in two steps as mentioned earlier. The step for logical deletion of "curr" is the \lp for the remove method.\\
After curr has been successfully deleted, it is added to the “Pool”. These deleted nodes can now be reused for later add operations.
\par
Now, what happens if any of the two CAS operations fail.\\
\textbf{Case 1:} CAS for logical deletion of curr fails. It implies that some other thread has performed a concurrent operation on curr and a synchronization conflict is detected. The current thread has to restart it's operation.\\
\textbf{Case 2:} CAS for physical deletion fails. It implies that  some other thread has performed a concurrent operation on pred. The current thread has two choices: it can depend on other traversing threads to “help” physically delete curr or it can traverse the list once more time to ensure curr's deletion.\\
An important note is that incrementing the pred's stamp by 2 during physical deletion prevents the ABA problem.

\begin{figure}[h!]
\centering
\includegraphics[scale=0.25]{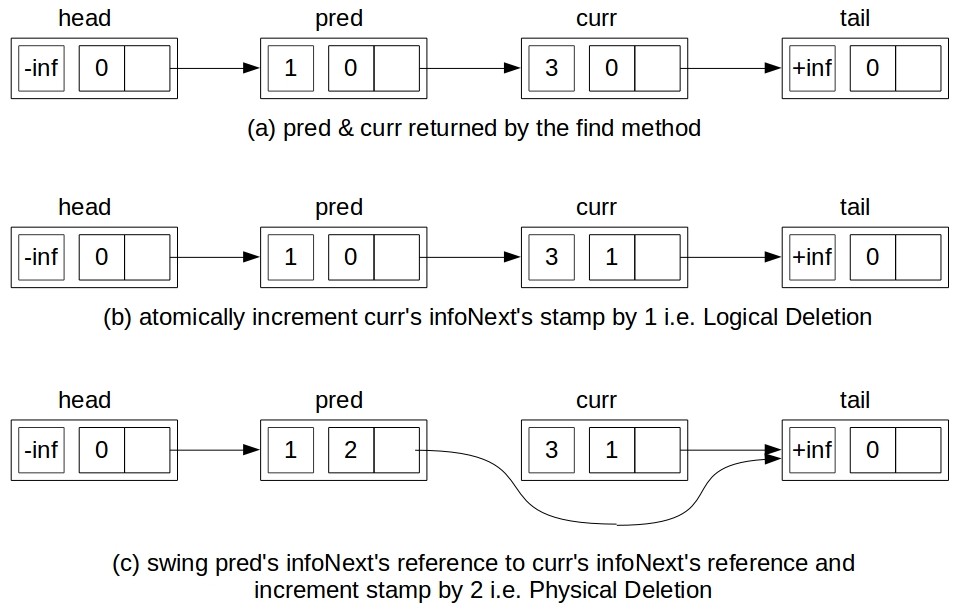}
\caption{GCLFList: Remove Steps}
\end{figure}

\begin{figure}[h!]
\centering
\includegraphics[scale=0.24]{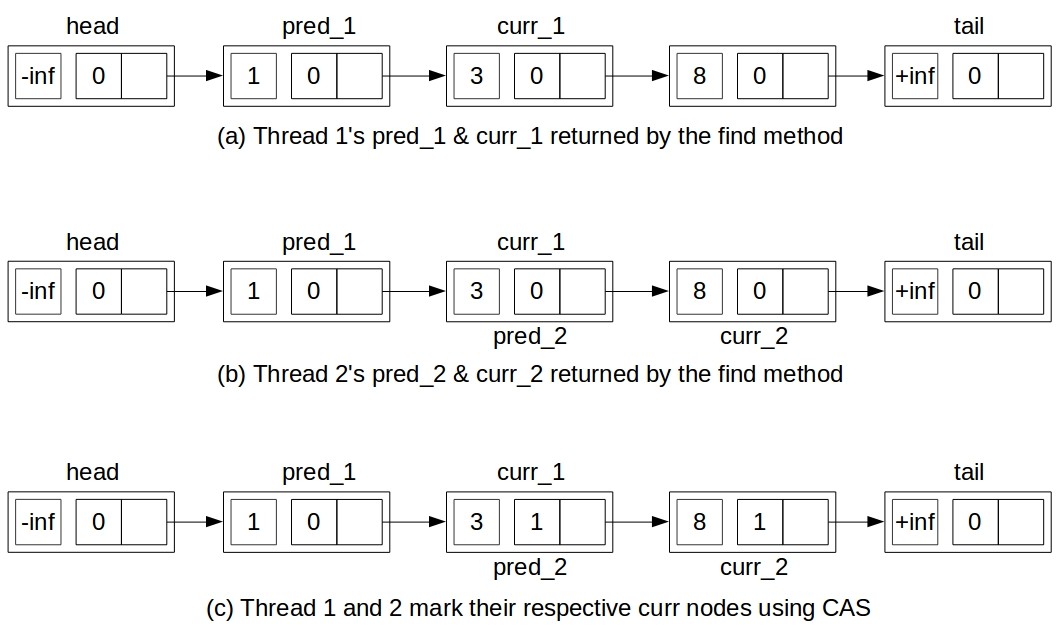}
\caption{GCLFList: Two concurrent removing Threads(Part 1)}
\end{figure}

\begin{figure}[h!]
\centering
\includegraphics[scale=0.24]{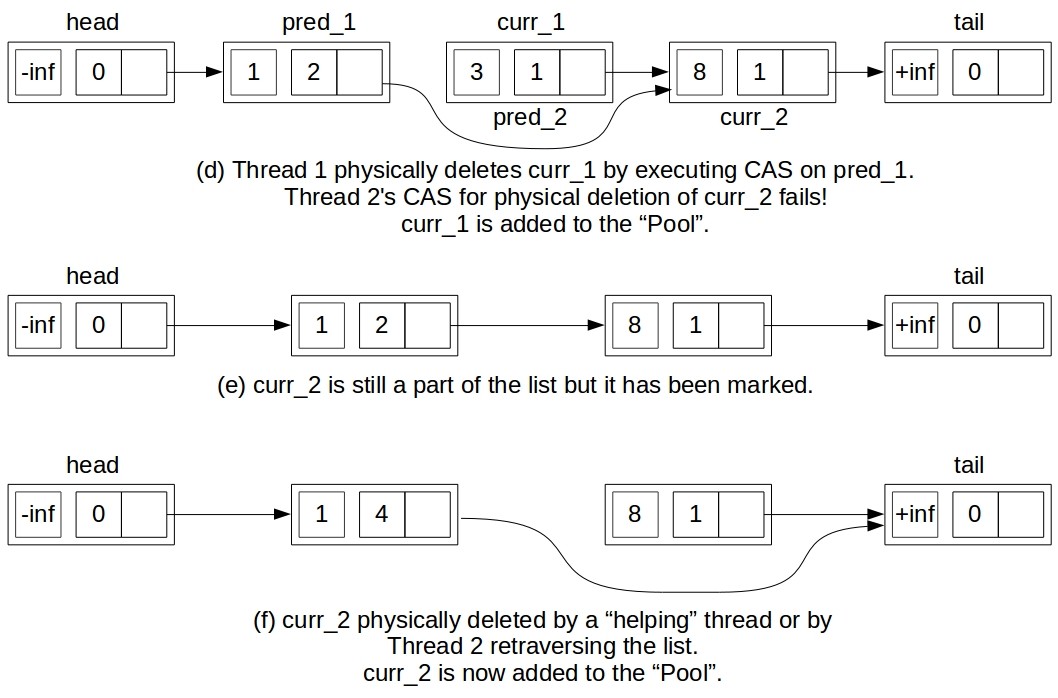}
\caption{GCLFList: Two concurrent removing Threads(Part 2)}
\end{figure}

\subsubsection{The add method}
The add method is used to add a key to the list if and only if the key is not already present in the list. It calls the “find” method to determine the correct pair of nodes for the add operation. The thread then queries the Pool for a node. If the Pool is not empty, a node is returned to be reused. Else, the thread creates a new node. It then inserts the new node, unlocks pred and curr and returns true. If the node is obtained from the Pool, it's stamp is incremented by 1 before inserting it into the list.
\par
An important observation to be made her is if the adding thread's CAS call on pred to insert the new node to the list fails, it calls the find method again, resulting in a new pair of pred and curr. However, another concurrent adding thread may have meanwhile added the same key to the list. The current thread now cannot add the same key anymore and has to return false. Before doing that, if the node was retrieved from the Pool, it is added back again to it. Else, if it was a newly created node, we can delete it since we have a guarantee that no other thread has a reference to it.
\par
This scenario never occured in GCLBList since once pred and curr were locked and validated and the key was previously absent from the list, there is a guarantee that the current thread would be able to add the key to the list successfully. Provided it doesn't crash midway before that.\\
The CAS call to set pred's infoNext's reference to the new node is the \textbf{\lp} for this method.

\begin{figure}[h!]
\centering
\includegraphics[scale=0.24]{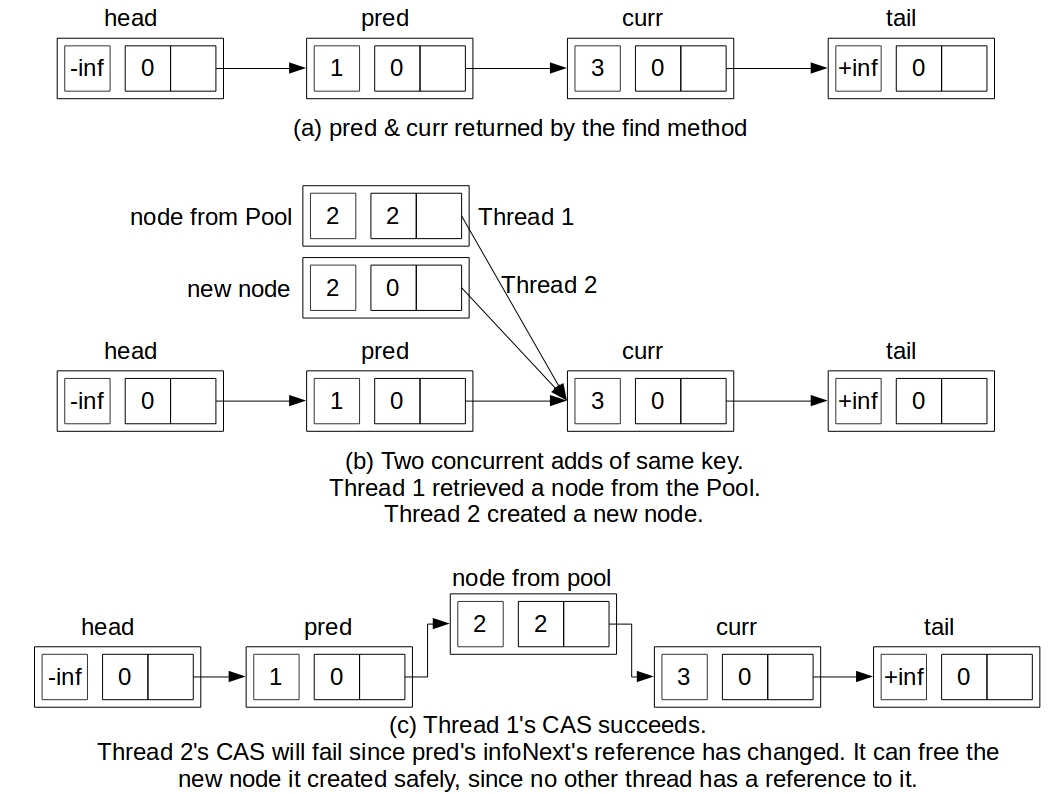}
\caption{GCLFList: Add Steps}
\end{figure}

\subsubsection{The contains method}
The contains method starts from the “head” node and keeps traversing the list in an optimistic hand-over-hand fashion. At every step of the traversal, the next reference and stamp of a node is read atomically\cite{ASR}. The thread keeps traversing the list until it finds the first node with a key greater than or equal to the key that is being searched.
\par
Again, stamps are used to detect synchronization conflicts during traversal. If at any time during a thread's traversal, the stamp of the pred node changes, a synchronization conflict with another “removing” thread is detected. The current thread “retries” it's traversal from the head node.\\
The method returns true if and only if the key is present in the list and it's infoNext's stamp is even. A successful contains is \textbf{linearized} when a matching key is found and the it's stamp is even.

%% file: gc-pool.tex
\section{The Pool}
The pool is a concurrent data structure that is used to hold the deleted nodes that have been reclaimed from the list. Ideally, any data structure that treats the node object as a "payload" can be used as the pool. In our experiments we used two different queue implementations to act as the pool. The code for both the queues has been kept in the appendix.
\subsection{The blocking unbounded total queue}
This lock-based concurrent queue\cite{HerlihyShavit:AMP:Book:2012} uses two separate locks for each queue operation i.e. an enqLock to enqueue a deleted node to the queue and a deqLock to dequeue a node from the queue, respectively.
\par
Before a thread performs an enqueue or a dequeue operation, it acquires the corresponding lock on the queue. After acquiring the lock, the thread performs it's operation and releases the lock upon completion. The lock ensures that, at a particular time, only one thread is able to perform an enqueue or a dequeue operation on the queue.

\subsection{The unbounded Lock-free queue}
This lock-free concurrent queue\cite{HerlihyShavit:AMP:Book:2012} uses atomic compareAndSet or CAS calls instead of locks for the queue operations. The CAS calls are used to enqueue a node into the queue and also to dequeue a node from the queue.
\par
This lock-free implementation helps to prevent faster threads from starving, with the removal of coarse-grained locks. This queue implementation also uses the concept of helping, where faster thread help the slower threads to finish their queue operations.
\par
The enqueue operation is done in two steps:\\
- The thread locates the last node in the queue and uses a CAS call to append the new node to the queue.\\
- It then uses another CAS call to change the queue's tail from the previous last node to the current last node.

\par
Since the above two CAS calls are not a single atomic operation, threads help each other to complete the second CAS, if a half finished enqueue operation is encountered.

\par
An important attribute to be noted about the queue is that it also uses the AtomicStampedReference\cite{ASR} object, in its' head and tail, to prevent the ABA problem\cite{queue_aba} problem from occurring in the queue.

%% file: evaluation.tex
\section{Evaluation}
We tested both versions of GCList against existing implementations of a concurrent set namely: LazyList\cite{lazy_list}, Hand-over-Hand List\cite{hoh}, Harris's LockFreeList\cite{lfree}, a Shared Pointer\cite{SPtr} version of LazyList and a LockFreeList using Hazard Pointers.
\par
We used both the lock-based queue and the lock-free queue, as a Pool, in combination with the two versions of GCList. The resultant set representation is named by using the list's name as prefix and pool's name as suffix. For example, the GCLBList using the lock-based queue would be named GCLBListLBQueue and the GCLFList using the lock-free queue would be named GCLFListLFQueue.\\
The LazyList based on Shared Pointers has been named LazyList\_SP. The LockFreeList using Hazard Pointers for Memory Reclamation has been named LockFreeList\_HP.
\par
We tested the above mentioned algorithms versus our algorithms for both performance and memory consumption, with varying workloads and randomized Set operations.
\par
For performance, we fix the total number of operations that each thread can perform, divided in varying ratios between adds, removes and contains. We allowed each algorithm to run for 10 seconds and measured the number of operations completed during said time period.\\
For memory consumption, we fix the total number of operations that each thread can perform, divided in varying ratios between adds, removes and contains. We keep track of the number of times each thread allocates and deallocates memory. Whenever the thread allocates new memory, a thread-local variable is incremented and whenever the memory is released, the variable is decremented. At the end of all thread operations, the main thread consolidates the sum of all the thread-local variables. We take the ratio of a List's node count versus the Hand-Over-Hand List. We use this ratio to compare the memory consumed by an algorithm during it's entire execution.
\par
Based on the above criteria, we obtained the following graphs.

\input{graphs}

\subsection{Analysis of Results}
\subsubsection{Performance Analysis}
From the graphs, we can see that the performance of both versions of GCList i.e GCLBList and GCLFList, is at par or even better than Harris's LockFreeList. Both outperform the Hand-over-Hand List, the LazyList based on Shared Pointers and the LockFreeList using Hazard Pointers for memory reclamation by multiple folds. The GCList versions are only outperformed by the original LazyList.

\subsubsection{Memory Consumption Analysis}
However, in terms of Memory consumption, both versions of GCList consume a lot less memory than the original LazyList. It also needs less memory than Harris's LockFreeList and the Hand-over-Hand List. In comparison with generic techniques like Shared Pointers and Hazard Pointers, memory consumption of GCList is still comparable to both.
\par
The plots for LazyList and LockFreeList have not been shown in the graphs. This is because they consume way too much memory compared to the other lists. Adding the plots for LazyList and LockFreeList reduces the other plots to straight lines similar to the Hand-over-Hand plot. This is due to the fact that LazyList and LockFreeList are unable to either free deleted nodes or reuse them. For each insert operation, new memory has to be allocated for the node.

%% file: graphs.tex
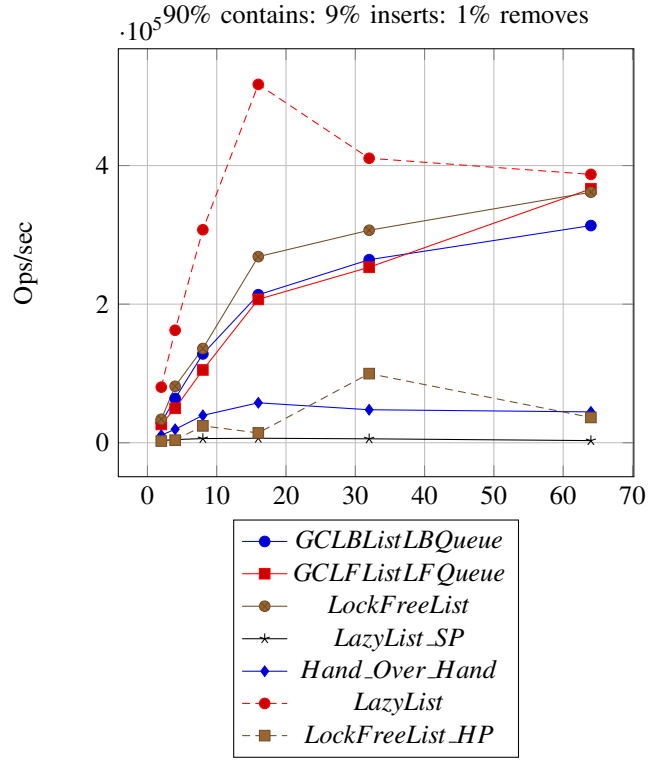
\begin{figure}
\begin{tikzpicture}
\begin{axis}[
title = 90\% contains: 9\% inserts: 1\% removes,
align=left,
xlabel={Number of Threads},
ylabel={Ops/sec},
grid=major,
legend entries={$GCLBListLBQueue$,$GCLFListLFQueue$,$LockFreeList$,$LazyList\_SP$,$Hand\_Over\_Hand$,$LazyList$,$LockFreeList\_HP$},
legend style={at={(0.5,-0.1)},anchor=north}
]
\addplot table [x=Threads, y=GCLBListLBQueue, col sep=comma] {Per_10.csv};
\addplot table [x=Threads, y=GCLFListLFQueue, col sep=comma] {Per_10.csv};
\addplot table [x=Threads, y=LockFreeList, col sep=comma] {Per_10.csv};
\addplot table [x=Threads, y=LazyList_SP, col sep=comma] {Per_10.csv};
\addplot table [x=Threads, y=Hand_Over_Hand, col sep=comma] {Per_10.csv};
\addplot table [x=Threads, y=LazyList, col sep=comma] {Per_10.csv};
\addplot table [x=Threads, y=LockFreeList_HP, col sep=comma] {Per_10.csv};
\end{axis}
\end{tikzpicture}
\caption{Performance Analysis with 10\% writes}
\end{figure}

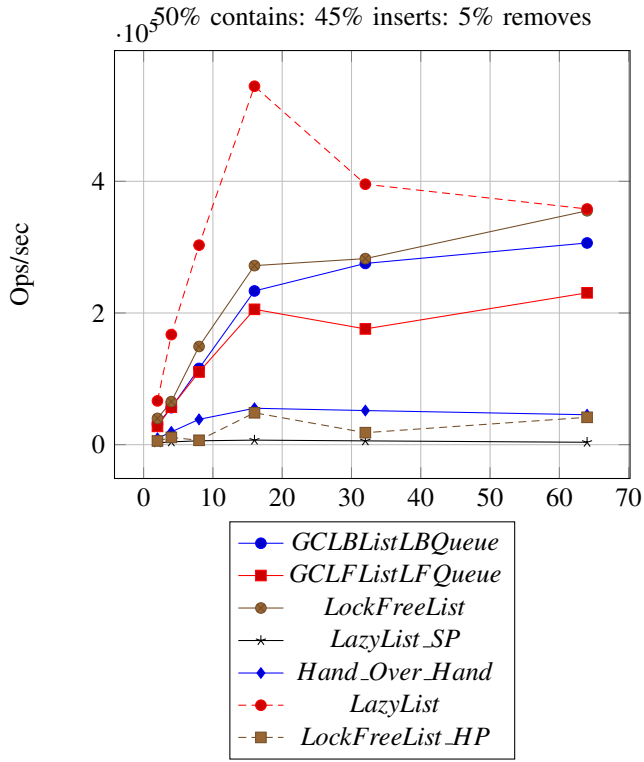
\begin{figure}
\begin{tikzpicture}
\begin{axis}[
title = 50\% contains: 45\% inserts: 5\% removes,
align=left,
xlabel={Number of Threads},
ylabel={Ops/sec},
grid=major,
legend entries={$GCLBListLBQueue$,$GCLFListLFQueue$,$LockFreeList$,$LazyList\_SP$,$Hand\_Over\_Hand$,$LazyList$,$LockFreeList\_HP$},
legend style={at={(0.5,-0.1)},anchor=north}
]
\addplot table [x=Threads, y=GCLBListLBQueue, col sep=comma] {Per_50.csv};
\addplot table [x=Threads, y=GCLFListLFQueue, col sep=comma] {Per_50.csv};
\addplot table [x=Threads, y=LockFreeList, col sep=comma] {Per_50.csv};
\addplot table [x=Threads, y=LazyList_SP, col sep=comma] {Per_50.csv};
\addplot table [x=Threads, y=Hand_Over_Hand, col sep=comma] {Per_50.csv};
\addplot table [x=Threads, y=LazyList, col sep=comma] {Per_50.csv};
\addplot table [x=Threads, y=LockFreeList_HP, col sep=comma] {Per_50.csv};
\end{axis}
\end{tikzpicture}
\caption{Performance Analysis with 50\% writes}
\end{figure}

\begin{figure}
\begin{tikzpicture}
\begin{axis}[
title = 30\% contains: 63\% inserts: 7\% removes,
align=left,
xlabel={Number of Threads},
ylabel={Ops/sec},
grid=major,
legend entries={$GCLBListLBQueue$,$GCLFListLFQueue$,$LockFreeList$,$LazyList\_SP$,$Hand\_Over\_Hand$,$LazyList$,$LockFreeList\_HP$},
legend style={at={(0.5,-0.1)},anchor=north}
]
\addplot table [x=Threads, y=GCLBListLBQueue, col sep=comma] {Per_70.csv};
\addplot table [x=Threads, y=GCLFListLFQueue, col sep=comma] {Per_70.csv};
\addplot table [x=Threads, y=LockFreeList, col sep=comma] {Per_70.csv};
\addplot table [x=Threads, y=LazyList_SP, col sep=comma] {Per_70.csv};
\addplot table [x=Threads, y=Hand_Over_Hand, col sep=comma] {Per_70.csv};
\addplot table [x=Threads, y=LazyList, col sep=comma] {Per_70.csv};
\addplot table [x=Threads, y=LockFreeList_HP, col sep=comma] {Per_70.csv};
\end{axis}
\end{tikzpicture}
\caption{Performance Analysis with 70\% writes}
\end{figure}

\begin{figure}
\begin{tikzpicture}
\begin{axis}[
title = 90\% contains: 9\% inserts: 1\% removes,
align=left,
xlabel={Number of Threads},
ylabel={Node Count ratio vs Hand-Over-Hand List},
grid=major,
legend entries={$GCLBListLBQueue$,$GCLFListLFQueue$,$LazyList\_SP$,$Hand\_Over\_Hand$,$LockFreeList\_HP$},
legend style={at={(0.5,-0.1)},anchor=north}
]
\addplot table [x=Threads, y=GCLBListLBQueue, col sep=comma] {Mem_10.csv};
\addplot table [x=Threads, y=GCLFListLFQueue, col sep=comma] {Mem_10.csv};
\addplot table [x=Threads, y=LazyList_SP, col sep=comma] {Mem_10.csv};
\addplot table [x=Threads, y=Hand_Over_Hand, col sep=comma] {Mem_10.csv};
\addplot table [x=Threads, y=LockFreeList_HP, col sep=comma] {Mem_10.csv};
\end{axis}
\end{tikzpicture}
\caption{Memory Consumption Analysis with 10\% writes}
\end{figure}

\begin{figure}
\begin{tikzpicture}
\begin{axis}[
title = 50\% contains: 45\% inserts: 5\% removes,
align=left,
xlabel={Number of Threads},
ylabel={Node Count ratio vs Hand-Over-Hand List},
grid=major,
legend entries={$GCLBListLBQueue$,$GCLFListLFQueue$,$LazyList\_SP$,$Hand\_Over\_Hand$,$LockFreeList\_HP$},
legend style={at={(0.5,-0.1)},anchor=north}
]
\addplot table [x=Threads, y=GCLBListLBQueue, col sep=comma] {Mem_50.csv};
\addplot table [x=Threads, y=GCLFListLFQueue, col sep=comma] {Mem_50.csv};
\addplot table [x=Threads, y=LazyList_SP, col sep=comma] {Mem_50.csv};
\addplot table [x=Threads, y=Hand_Over_Hand, col sep=comma] {Mem_50.csv};
\addplot table [x=Threads, y=LockFreeList_HP, col sep=comma] {Mem_50.csv};
\end{axis}
\end{tikzpicture}
\caption{Memory Consumption Analysis with 50\% writes}
\end{figure}
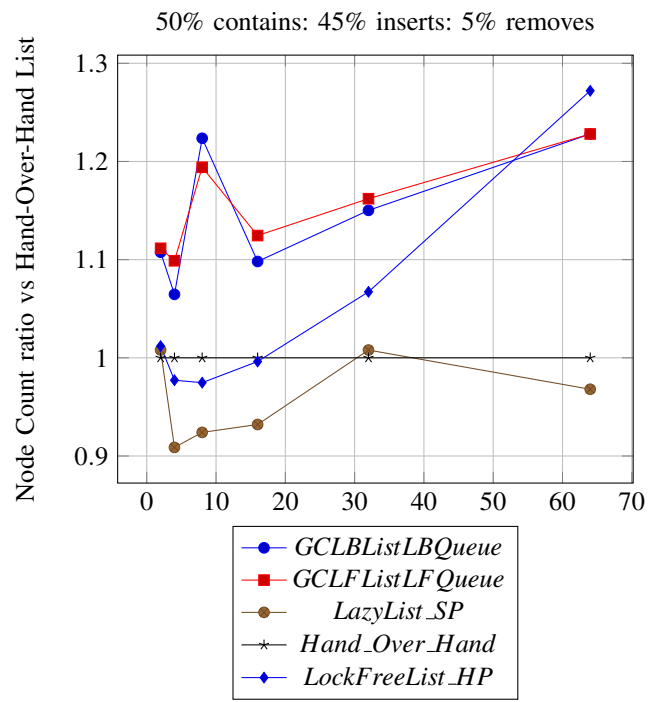

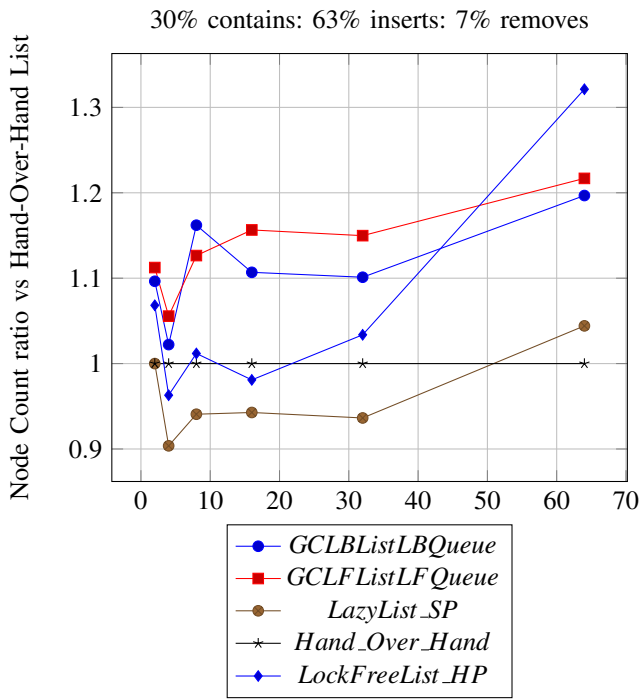
\begin{figure}
\begin{tikzpicture}
\begin{axis}[
title = 30\% contains: 63\% inserts: 7\% removes,
align=left,
xlabel={Number of Threads},
ylabel={Node Count ratio vs Hand-Over-Hand List},
grid=major,
legend entries={$GCLBListLBQueue$,$GCLFListLFQueue$,$LazyList\_SP$,$Hand\_Over\_Hand$,$LockFreeList\_HP$},
legend style={at={(0.5,-0.1)},anchor=north}
]
\addplot table [x=Threads, y=GCLBListLBQueue, col sep=comma] {Mem_70.csv};
\addplot table [x=Threads, y=GCLFListLFQueue, col sep=comma] {Mem_70.csv};
\addplot table [x=Threads, y=LazyList_SP, col sep=comma] {Mem_70.csv};
\addplot table [x=Threads, y=Hand_Over_Hand, col sep=comma] {Mem_70.csv};
\addplot table [x=Threads, y=LockFreeList_HP, col sep=comma] {Mem_70.csv};
\end{axis}
\end{tikzpicture}
\caption{Memory Consumption Analysis with 70\% writes}
\end{figure}

%% file: conc.tex
\section{Conclusion and Future Work}
In this paper, we have presented \textbf{GCList}, a linked-list representation of a concurrent set, with in-built garbage collection. Both the lock-based and lock-free versions of \textbf{GCList}, i.e. \textbf{GCLBList} and \textbf{GCLFList}, are introduced.
\par
Our results show that \textbf{GCList} matches or outperforms most of the existing representations of a concurrent set, while consuming a lot lesser memory than the higher-performing algorithms like LazyList. Memory consumption was at par with generic garbage collection facilities like Shared Pointers and Hazard Pointers, while outperforming them many folds.
\par
In future work, we plan to investigate whether we can extend it to other data structures similar to a concurrent list or using it as a part of it's structure e.g. SkipList, Hash Tables etc.

%% file: appendix.tex
\onecolumn
\clearpage

\section*{Appendices}

\subsection{GCLBList Pseudo Code}

\begin{breakablealgorithm}
\caption{The find method}\label{GCLBList_find}
\begin{algorithmic}[1]
\Function{Window find}{$Node\ head,\ int\ key$}\Comment{Traverse from head and find node with key-value 'key'}
\If{$head.infoNext.getReference()\ ==\ tail$}\Comment{\parbox[t]{.4\linewidth}{head \& tail are the only nodes in the list}}
	\State \Return {$Window\ (head,\ tail,\ head.infoNext.getStamp()\ ,\ tail.infoNext.getStamp())$}
\EndIf
\While{$true$}\label{retry}
	\State $pred\gets head$ \Comment{Start from the head}
    \State $curr \gets pred.infoNext.get(predSt)$ \Comment{Read pred’s infoNext’s reference \& stamp atomically}
    \While{$true$}
    	\State $breakTest \gets key\le curr.key$ \Comment{\parbox[t]{.5\linewidth}{Break when key-value greater than or equal to required key is found}}
        \State $succ \gets curr.infoNext.get(currSt)$
        \Comment{\parbox[t]{.45\linewidth}{Read curr’s infoNext’s reference \& stamp atomically. succ may be null if curr has been deleted}}
        \State $nPredSt \gets pred.infoNext.getStamp()$
        \Statex \Comment{\parbox[t]{.45\linewidth}{Read pred’s stamp again before advancing forward. This is the safety check to ensure we are traversing the list correctly, in increasing order of keys}}
        \If{$predSt \neq nPredSt$} 
        	\State{\textbf{go to 5}} 
            \Comment{\parbox[t]{.6\linewidth}{If pred’s new stamp is different from the one read previously, a synchronization conflict is detected. curr may have been deleted by another thread from the list. The thread restarts it’s traversal to ensure correctness. If pred’s stamp is still the same, then everything is fine.}}
        \EndIf
        \If{$breakTest$}
        \State{\textbf{go to 22}}%
        \Comment{\parbox[t]{.67\linewidth}{If pred’s stamp has not changed, everything is fine. Check if required pair of nodes has been found. If yes, break. Else, continue.}}
        \EndIf
        \State $pred \gets curr$ \Comment{Keep advancing pred and curr in the list}
        \State $curr \gets succ$
        \State $predSt \gets currSt$ \Comment{\parbox[t]{.65\linewidth}{Keep track of new pred’s old stamp to be used later, to detect synchronization conflicts}}
    \EndWhile
    \State \Return {$Window(pred,\ curr,\ predSt,\ currSt)$} \label{return}\Comment{\parbox[t]{.4\linewidth}{Return pred and curr, along with their stamps, encapsulated in a window object.}}
\EndWhile
\EndFunction
\end{algorithmic}
\end{breakablealgorithm}

\begin{algorithm}
\caption{The validate method}\label{GCLBList_validate}
\begin{algorithmic}[1]
\Function{bool validate}{$Node\ pred,\ int\ predSt,\ Node\ curr,\ int\ currSt$}
\Statex \Comment{\parbox[t]{.56\linewidth}{Checks consistency of locked nodes 'pred' \& 'curr', using their stamps, predSt \& currSt}}
	\State $nCurr \gets pred.infoNext.get(predSt)$ \Comment{Re-read pred’s infoNext’s reference and stamp atomically}
    \State $nCurrSt \gets curr.infoNext.getStamp()$ \Comment{Re-read curr’s infoNext’s stamp atomically}
    \State \Return {$predSt\ ==\ nPredSt\ \&\&\ currSt\ ==\ nCurrSt\ \&\&\ curr\ ==\ nCurr$}
    \Statex \Comment{\parbox[t]{.83\linewidth}{Checks if pred is still pointing to curr. And if any of their stamps have changed from their old values. If yes, a conflict is detected. Returns true or false to calling method.}}
\EndFunction
\end{algorithmic}
\end{algorithm}

\begin{algorithm}[H]
\caption{The remove method}\label{GCLBList_remove}
\begin{algorithmic}[1]
\Function{bool remove}{$Node\ head,\ int\ key$}\Comment{\parbox[t]{.5\linewidth}{Remove a node with key-value 'key' from the list}}
	\While{$true$} \label{retry}
    	\State $window \gets find(head, key)$
        \State $pred \gets window.pred, curr \gets window.curr$
        \State $predSt \gets window.predSt, currSt \gets window.currSt$
        \Statex \Comment{Retrieve pred and curr, and their stamps, from the window object}
        \State{$pred.lock()$}
        \If{$!curr.tryLock()$}
        	\State $pred.unlock()$
            \State{\textbf{go to 6}} 
            \Comment{\parbox[t]{.7\linewidth}{Lock both the nodes. tryLock() is to prevent deadlocks, since there is no guarantee, that keys are being locked in increasing order}}
        \EndIf
        \If{$validate(pred,\ predSt,\ curr,\ currSt)$} \Comment{\parbox[t]{.45\linewidth}{Use validate to ensure the consistency of pred and curr}}
        	\If{$curr.key \neq key$}
            	\State $curr.unlock()$
                \State $pred.unlock()$
                \State \Return $false$
                \Comment{If key is not present, unlock both nodes. And return false}
            \Else
            	\State $stamp \gets pred.infoNext.getStamp()$
                \State $temp \gets curr.infoNext.getReference()$
                \State $pred.infoNext.set(temp,++stamp)$
                \Comment{\parbox[t]{.4\linewidth}{Deletion Step 1: atomically swing pred’s infoNext’s reference to curr’s infoNext’s reference and increment pred’s infoNext’s stamp by 1}}
                
                \State $temp \gets curr.infoNext.get(stamp)$
                \State $curr.infoNext.set(temp,++stamp)$
                \Comment{\parbox[t]{.4\linewidth}{Deletion Step 2: atomically increment curr’s infoNext’s stamp by 1}}
                
                \State $Pool.set(curr)$
                \Comment{\parbox[t]{.6\linewidth}{Add deleted node 'curr' to the Pool. curr can be reused for later add operations}}
                
                \State $curr.unlock()$
                \State $pred.unlock()$
                \State \Return $true$
                \Comment{\parbox[t]{.6\linewidth}{Unlock pred and curr. Return true}}
            \EndIf
        \EndIf
        \State $curr.unlock()$
        \State $pred.unlock()$
    \EndWhile
\EndFunction
\end{algorithmic}
\end{algorithm}

\begin{algorithm}
\caption{The add method}\label{GCLBList_add}
\begin{algorithmic}[1]
\Function{bool add}{$Node\ head,\ int\ key$}\Comment{Add a node with key-value 'key' from the list}
	\While{$true$} \label{retry}
    	\State $window \gets find(head, key)$
        \State $pred \gets window.pred, curr \gets window.curr$
        \State $predSt \gets window.predSt, currSt \gets window.currSt$
        \Statex \Comment{Retrieve pred and curr, and their stamps, from the window object}
        \State{$pred.lock()$}
        \If{$!curr.tryLock()$}
        	\State {$pred.unlock()$}
            \State{\textbf{go to 6}} 
            \Comment{\parbox[t]{.7\linewidth}{Lock both the nodes. tryLock() is to prevent deadlocks, since there is no guarantee, that keys are being locked in increasing order}}
        \EndIf
        \If{$validate(pred, predSt, curr, currSt)$} \Comment{\parbox[t]{.45\linewidth}{Use validate to ensure the consistency of pred and curr}}
        	\If{$curr.key == key$}
            	\State $curr.unlock()$
                \State $pred.unlock()$
                \State \Return $false$
                \Comment{If key is already present, unlock both nodes. And return false}
            \Else
            	\State $node \gets Pool.get()$
                \Comment{Query the Pool for a node}
                
                \If{$node \neq nullptr$}
                	\State $node.key \gets key$
                    \Comment{node has been retrieved from pool. Reuse for new add operation}
                \Else
                	\State $node \gets new Node(key)$
                    \Comment{Pool is empty. Create new node.}
                \EndIf
                
                \State $stamp \gets node.infoNext.getStamp()$
                \State $node.infoNext.set(curr, stamp)$
                \Comment{\parbox[t]{.4\linewidth}{Set new node’s reference to curr. No need to change new node’s stamp}}
                
                \State $stamp \gets pred.infoNext.getStamp()$
                \State $pred.infoNext.set(node, stamp)$
                \Comment{\parbox[t]{.4\linewidth}{Atomically set pred’s infoNext’s reference to new node. No need to change pred’s stamp}}
                
                \State $curr.unlock()$
                \State $pred.unlock()$
                \State \Return $true$
                \Comment{\parbox[t]{.4\linewidth}{Unlock pred and curr. Return true}}
            \EndIf
        \EndIf
        \State $curr.unlock()$
        \State $pred.unlock()$
    \EndWhile
\EndFunction
\end{algorithmic}
\end{algorithm}

\begin{algorithm}
\caption{The contains method}\label{GCLBList_contains}
\begin{algorithmic}[1]
\Function{bool contains}{$Node\ head,\ int\ key$}\Comment{Traverse from head and find node with key-value 'key'}
\State $breakTest \gets false$
\While{true}\label{retry}
	\State $pred\gets head$ \Comment{Start from the head}
    \State $curr \gets pred.infoNext.get(predSt)$
    \Comment{Read pred’s infoNext’s reference \& stamp atomically}
    \State $currKey \gets curr.key$
    \Comment{Read curr's key-value}
    
    \While{$true$}
    	\State $breakTest \gets key\le currKey$
        \Comment{\parbox[t]{.45\linewidth}{Break when key-value greater than or equal to required key is found}}
        \State $succ \gets curr.infoNext.get(currSt)$
        \Comment{\parbox[t]{.45\linewidth}{Read curr’s infoNext’s reference \& stamp atomically. succ may be null if curr has been deleted}}
        
        \State $nPredSt \gets pred.infoNext.getStamp()$
        \Statex \Comment{\parbox[t]{.45\linewidth}{Read pred’s stamp again before advancing forward. This is the safety check to ensure we are traversing the list correctly, in increasing order of keys}}
        \If{$predSt \neq nPredSt$} 
        	\State{\textbf{go to 3}} 
            \Comment{\parbox[t]{.6\linewidth}{If pred’s new stamp is different from the one read previously, a synchronization conflict is detected. curr may have been deleted by another thread from the list. The thread restarts it’s traversal to ensure correctness. If pred’s stamp is still the same, then everything is fine.}}
        \EndIf
        \If{$breakTest$}
        	\State{\textbf{go to 22}} 
        	\Comment{\parbox[t]{.6\linewidth}{If pred’s stamp has not changed, everything is fine. Check if required pair of nodes has been found. If yes, break. Else, continue.}}
        \EndIf
        
        \State $pred \gets curr$
        \Comment{Keep advancing pred and curr in the list}
        \State $curr \gets succ$
        \State $predSt \gets currSt$
        \Comment{\parbox[t]{.6\linewidth}{Keep track of new pred’s old stamp to be used later, to detect synchronization conflicts}}
        
        \State $currKey \gets curr.key$
        \Comment{Read curr's key-value}
    \EndWhile
    \State \Return {$currKey == key$} \label{return}
    \Comment{\parbox[t]{.6\linewidth}{Return true if 'key' has been found. Else, return false.}}
\EndWhile
\EndFunction
\end{algorithmic}
\end{algorithm}

\clearpage
\subsection{GCLFList Pseudo Code}

\begin{breakablealgorithm}
\caption{The find method}\label{GCLFList_find}
\begin{algorithmic}[1]
\Function{Window find}{$Node\ head,\ int key,\ Node\ prevCurr$}
\Statex \Comment{\parbox[t]{.4\linewidth}{Traverse from head and find node with key-value 'key'}}
	\State $breakTest \gets false, snip \gets false$
    \While{$true$}\label{retry}
    	\State $pred \gets head$
        \Comment{Start from the head}
        \State $curr \gets pred.infoNext.get(predSt)$
        \Comment{Read curr’s infoNext’s reference \& stamp atomically}
        
        \While{true}
        	\State $currKey \gets curr.key$
            \Comment{Read curr’s key value}
            \State $succ \gets curr.infoNext.get(currSt)$
            \Comment{\parbox[t]{.45\linewidth}{Atomically read curr’s infoNext’s reference and stamp. Successor may be null if curr has been deleted}}
            
            \If{$currSt\mod 2 ==1$}
            	\State $snip \gets pred.infoNext.compareAndSet(curr, succ, predSt, predSt+2)$
                \Statex \Comment{\parbox[t]{.67\linewidth}{This is the "\textbf{helping}" step. If curr is marked(stamp is odd), attempt to physically remove from the list. Done by calling an atomic CAS operation on pred, to atomically set pred’s infoNext’s reference to successor and increment stamp by 2}}
                \If{$!snip$}
                	\State{\textbf{go to 3}} 
                    \Comment{If the CAS operation fails, restart the traversal}
                \EndIf
                
                \State $Pool.set(curr)$
                \Comment{Else, add curr to the Pool.}
                \State $predSt+=2$
                \Comment{And keep track of updated pred’s stamp}
                
            \EndIf
            
            \State $breakTest \gets key \le currKey$
            \Comment{\parbox[t]{.45\linewidth}{Break when key greater than or equal to required key is found}}
            
            \State $nPredSt \gets pred.infoNext.getStamp()$
            \Statex \Comment{\parbox[t]{.45\linewidth}{Read pred’s stamp again before advancing forward. This is the safety check to ensure we are traversing the list correctly, in increasing order of keys}}
            
            \If{$predSt \neq nPredSt$}
            	\State\textbf{go to 3} 
                \Comment{\parbox[t]{.6\linewidth}{If pred’s new stamp is different from the one read previously, a synchronization conflict is detected. curr may have been deleted by another thread from the list. The thread restarts it’s traversal to ensure correctness. If pred’s stamp is still the same, then everything is fine}}
            \EndIf
            
            \If{$breakTest$}
            	\State{\textbf{go to 34}} 
                \Comment{\parbox[t]{.68\linewidth}{If pred’s stamp has not changed, everything is fine. Check if required pair of nodes has been found. If yes, break. Else, continue}}
            \EndIf
            
            \If{$!snip$}
            	\State{$pred \gets curr$}
                 \Comment{\parbox[t]{.63\linewidth}{If no helping was done i.e. no marked node was found,}}
                \State{$curr \gets succ$}         
                \Comment{\parbox[t]{.63\linewidth}{Keep advancing pred and curr in the list}}
                \State{$predSt \gets currSt$}
                \Comment{\parbox[t]{.63\linewidth}{Keep track of new pred’s old stamp to be used later, to detect synchronization conflicts}}
            \Else
            	\State{$curr \gets succ$}
                \Comment{\parbox[t]{.63\linewidth}{If helping was done to remove an encountered marked done, It implies pred is still the same. Advance only curr}}
                \State{$snip \gets false$}
            \EndIf
        \EndWhile
        \State \Return {$Window(pred,\ curr,\ predSt,\ currSt)$} \label{return}
        \Statex \Comment{\parbox[t]{.83\linewidth}{Return pred and curr, along with their stamps, encapsulated in a Window object}}
    \EndWhile
\EndFunction
\end{algorithmic}
\end{breakablealgorithm}

\begin{algorithm}
\caption{The remove method}\label{GCLFList_remove}
\begin{algorithmic}[1]
\Function{bool remove}{$Node\ head,\ int\ key$}\Comment{\parbox[t]{.45\linewidth}{Remove a node with key-value 'key' from the list}}
	\While{$true$} \label{rem_retry}
    	\State $window \gets find(head, key, nullptr)$
        \State $pred \gets window.pred, curr \gets window.curr$
        \State $predSt \gets window.predSt, currSt \gets window.currSt$
        \Statex \Comment{Retrieve pred and curr, and their stamps, from the window object}
        \If{$curr.key \neq key$}
        	\State \Return {$false$}
            \Comment{If key is not present, return false}
        \Else
        	\State{$succ \gets curr.infoNext.getReference()$}
            \Comment{Read curr’s infoNext’s reference}
            \State{$snip \gets curr.infoNext.compareAndSet(succ, succ, currSt, currSt+1)$}
            \Statex \Comment{\parbox[t]{.52\linewidth}{Deletion Step 1: Atomically increment curr’s infoNext’s stamp by 1 using CAS i.e. Logical Deletion}}
            \If{$!snip$}
            	\State{\textbf{go to 2}} 
                \Comment{\parbox[t]{.52\linewidth}{If CAS fails, restart the operation}}
            \EndIf
            
            \If{$pred.infoNext.compareAndSet(curr, succ, predSt, predSt+2)$}
            \Statex \Comment{\parbox[t]{.52\linewidth}{Deletion Step 2: Atomically swing pred’s infoNext’s reference to successor. And increment pred’s infoNext’s stamp by 2 i.e. Physical Deletion}}
            	\State{$Pool.set(curr)$}
                \Comment{\parbox[t]{.52\linewidth}{If physical deletion is successful, add curr to the Pool}}
            \Else
            	\State{$find(head, key, nullptr)$}
                \Comment{\parbox[t]{.52\linewidth}{This step is optional. If physical deletion is unsuccessful, retraverse the list to remove it. Or depend on some other thread to "help out"}}
            \EndIf
            
            \State \Return {$true$}
            \Comment{\parbox[t]{.7\linewidth}{Return true on successful deletion. \textbf{Note:} Will return true even if only Logical deletion is successful}}
        \EndIf
    \EndWhile
\EndFunction
\end{algorithmic}
\end{algorithm}

\begin{algorithm}
\caption{The add method}\label{GCLFList_add}
\begin{algorithmic}[1]
\Function{bool add}{$Node\ head,\ int\ key$}\Comment{add a node with key-value 'key' from the list}
	\State{$fromPool \gets false$}
    \State {$node \gets Pool.get()$}
    \Comment{Query the Pool for a node}
    
    \If{$node == nullptr$}
    	\State {$node \gets new Node(key)$}
        \Comment{If Pool is empty, create a new node}
        \State{$fromPool \gets false$}
    \Else
    	\State {$node.key \gets key$}
        \Comment{\parbox[t]{.5\linewidth}{Else, node successfully retrieved from the Pool.}}
        \State {$nodeSt \gets node.infoNext.getStamp()$}
        \State {$node.infoNext.set(nullptr, nodeSt+1)$}
        \Comment{\parbox[t]{.5\linewidth}{Increment new node’s stamp by 1, to make the stamp even}}
        \State {$fromPool \gets true$}
    \EndIf
    
    \While{$true$} \label{add_retry}
    	\State $window \gets find(head, key, nullptr)$
        \State $pred \gets window.pred, curr \gets window.curr$
        \State $predSt \gets window.predSt, currSt \gets window.currSt$
        \Statex \Comment{Retrieve pred and curr, and their stamps, from the window object}
        \If{$curr.key \neq key$}
        	\State {$nodeSt \gets node.infoNext.getStamp()$}
            \State {$node.infoNext.set(curr, nodeSt)$}
            \Comment{\parbox[t]{.34\linewidth}{If 'key' is not already present in the list, set new node’s infoNext’s reference to curr}}
            
            \If{$pred.infoNext.compareAndSet(curr, node, predSt, predSt)$}
            \Statex \Comment{\parbox[t]{.65\linewidth}{Attempt to atomically CAS pred’s infoNext’s reference to new node. If CAS succeeds, return true}}
            	\State \Return {$true$}
            \Else
            	\State{\textbf{go to 13}}
                \Comment{\parbox[t]{.65\linewidth}{Else, restart the operation. \textbf{Note:} Next iteration, some other thread may have added the new key instead. If so, then this thread will return false}}
            \EndIf
        \Else
        \Comment{Key is already present in the list}
        	\If{$!fromPool$}
            	\State{$delete\ node$}
                \Comment{\parbox[t]{.65\linewidth}{new node was newly created by this thread. It can be safely freed, since no other thread has a reference to this node}}
            \Else
            	\State {$nodeSt \gets node.infoNext.getStamp()$}
                \State{$node.infoNext.set(nullptr, nodeSt-1)$}
                \State{$Pool.set(node)$}
                \Comment{\parbox[t]{.63\linewidth}{node was retrieved from the Pool. Decrement node’s stamp to make it odd again and add the node back to the Pool}}
            \EndIf
            \State \Return {$false$}
            \Comment{Return false since 'key' already present}
        \EndIf
    \EndWhile
\EndFunction
\end{algorithmic}
\end{algorithm}

\begin{algorithm}
\caption{The contains method}\label{GCLFList_contains}
\begin{algorithmic}[1]
\Function{bool contains}{$Node\ head,\ int\ key$}\Comment{Traverse from head and find node with key-value 'key'}
	\State $breakTest \gets false$
	\While{true}\label{con_retry}
		\State $pred\gets head$ \Comment{Start from the head}
        \While{$true$}
        	\State {$curr \gets pred.infoNext.get(predSt)$}
            \Comment{\parbox[t]{.45\linewidth}{Read curr’s infoNext’s reference \& stamp atomically}}
            
            \State {$currKey \gets curr.key$}
            \Comment{Read curr’s key value}
            
            \State {$succ \gets curr.infoNext.get(currSt)$}
            \Comment{\parbox[t]{.45\linewidth}{Atomically read curr’s infoNext’s reference and stamp. Successor may be null if curr has been deleted}}
            
            \State {$breakTest \gets key \le currKey$}
            \Comment{\parbox[t]{.45\linewidth}{Break when key greater than or equal to required key is found}}
            
            \State {$nPredSt \gets pred.infoNext.getStamp()$}
            \Comment{\parbox[t]{.45\linewidth}{Read pred’s stamp again before advancing forward. This is the safety check to ensure we are traversing the list correctly, in increasing order of keys}}
            
            \If{$predSt \neq nPredSt$}
            	\State{\textbf{go to 3}} 
                \Comment{\parbox[t]{.6\linewidth}{If pred’s new stamp is different from the one read previously, a synchronization conflict is detected. curr may have been deleted by another thread from the list. The thread restarts it’s traversal to ensure correctness. If pred’s stamp is still the same, then everything is fine}}
            \EndIf
            
            \If{$breakTest$}
            	\State\textbf{go to 20} 
                \Comment{\parbox[t]{.65\linewidth}{If pred’s stamp has not changed, everything is fine. Check if required node has been found. If yes, break. Else, continue}}
            \EndIf
            
            \State {$pred \gets curr$}
            \Comment{Keep advancing pred in the list}
            
            \State {$predSt \gets currSt$}
            \Comment{\parbox[t]{.65\linewidth}{Keep track of new pred’s old stamp to be used later, to detect synchronization conflicts}}
        \EndWhile
        
        \State {$marked \gets currSt\mod 2 ==1$} \label{return}
        \Comment{Check if curr is marked i.e. odd stamp}
        
        \State \Return {$currKey == key \ \&\&\  !marked$}
        \Comment{\parbox[t]{.45\linewidth}{Return true if and only if key is found and node is unmarked. Else, return false}}
	\EndWhile
\EndFunction
\end{algorithmic}
\end{algorithm}

\clearpage
\subsection{The Pool}
\subsubsection{Lock-Based Queue} C++ code
\begin{lstlisting}[language=C++]
class LBQueue
{
public:
    class QNode;
    mutex enqLock,deqLock;
    QNode *head,*tail;

    class QNode
    {
    public:
        Node *node;
        AtomicStampedReference<QNode> next;

        QNode()
        {
            node = nullptr;
            next.set(nullptr, 0);
        }

        QNode(Node *node)
        {
            this->node = node;
            next.set(nullptr, 0);
        }
    };

    LBQueue() {
        //QNode *sentinel = new QNode();
        head = new QNode();
        tail = new QNode();
        (tail->next).set(nullptr, 0);
        (head->next).set(tail, 0);
    }

    void set(Node *node) {
        enqLock.lock();
        if(node != nullptr) {
            QNode *qNode = new QNode(node);
            int stamp = (tail->next).getStamp();
            (tail->next).set(qNode, stamp);
            tail = qNode;
        }
        enqLock.unlock();
    }

    Node* get() {
        Node* result = nullptr;
        deqLock.lock();
        if ((head->next).getReference() != tail) {
            result = head->node;
            head = (head->next).getReference();
        }
        deqLock.unlock();
        return result;
    }
};
\end{lstlisting}

\subsubsection{Lock-Free Queue} C++ code
\begin{lstlisting}[language=C++]
class LFQueue {
public:
  class QNode;
  AtomicStampedReference<QNode> *head;
  AtomicStampedReference<QNode> *tail;

  class QNode {
	public:
	Node *node;
	AtomicStampedReference<QNode> next;

	QNode() {
	  node = nullptr;
	  next.set(nullptr, 0);
	}

	QNode(Node *node) {
	  this->node = node;
	  next.set(nullptr, 0);
	}
  };

  LFQueue() {
	QNode *sentinel = new QNode();
	head = new AtomicStampedReference<QNode>(sentinel, 0);
	tail = new AtomicStampedReference<QNode>(sentinel, 0);
  }

  void set(Node *node) {
	int lastStamp, nextStamp, stamp;

	if (node == nullptr)
	  return;
	QNode *x = new QNode(node);

	while (true) {
	  QNode *last = tail->get(&lastStamp);
	  QNode *next = (last->next).get(&nextStamp);

	  if (last == tail->get(&stamp) && stamp == lastStamp) {
		if (next == nullptr) {
		  if ((last->next).compareAndSet(next, x, nextStamp, nextStamp+1)) {
			tail->compareAndSet(last, x, lastStamp, lastStamp+1);
			return;
		  }
		}
		else {
		  tail->compareAndSet(last, next, lastStamp, lastStamp+1);
		}
	  }
	}
  }

  Node* get() {
	int lastStamp, firstStamp, nextStamp, stamp;

	while (true) {
	  QNode *first = head->get(&firstStamp);
	  QNode *last = tail->get(&lastStamp);
	  QNode *next = (first->next).get(&nextStamp);

	  if (first == head->get(&stamp) && stamp == firstStamp) {
		if (first == last) {
		  if (next == nullptr) {
			return nullptr;
		  }
		  tail->compareAndSet(last, next, lastStamp, lastStamp+1);
		}
		else {
		  Node *ret = first->node;
		  if (head->compareAndSet(first, next, firstStamp, firstStamp+1)) {
			return ret;
		  }
		}
	  }
	}
  }
};
\end{lstlisting}